# Giant spin-valve effect and chiral anomaly in antiferromagnetic topological insulators Mn(Bi$_{1-x}$Sb$_x$)$_2$Te$_4$

Seng Huat Lee[1,2], David Graf[3], Robert Robinson[2], John Singleton[4], Johanna C. Palmstrom[4], Zhiqiang Mao[1,2,5*]

[1]2D Crystal Consortium, Materials Research Institute, The Pennsylvania State University, University Park, Pennsylvania 16802, USA

[2]Department of Physics, The Pennsylvania State University, University Park, Pennsylvania 16802, USA

[3]National High Magnetic Field Lab, Tallahassee, Florida 32310, USA

[4]National High Magnetic Field Laboratory, Pulse Field Facility, Los Alamos National Laboratory, Los Alamos, New Mexico 87545, USA

[5]Department of Materials Science and Engineering, The Pennsylvania State University, University Park, PA 16802, USA

**We report *c*-axis transport studies on magnetic topological insulators Mn(Bi$_{1-x}$Sb$_x$)$_2$Te$_4$. We performed systematic *c*-axis magnetoresistivity measurements under high magnetic fields (up to 35 T) on several representative samples. We find the lightly hole- and lightly electron-doped samples, while both having the same order of magnitude of carrier density and similar spin-flop transitions, exhibit sharp contrast in electronic anisotropy and transport mechanism. The electronic anisotropy is remarkably enhanced for the lightly hole-doped sample relative to pristine MnBi$_2$Te$_4$ but not for the lightly electron-doped sample. The lightly electron-doped sample displays a giant negative longitudinal magnetoresistivity (LMR) induced by the spin-valve effect at the spin-flop transition field, whereas the lightly hole-doped sample exhibits remarkable negative LMR consistent with the chiral anomaly** **behavior of a Weyl semimetal. Furthermore, we find the large negative LMR of the lightly hole-doped sample extends to a wide temperature range above the Néel temperature ($T_N$) where the magnetoconductivity is proportional to $B^2$. This fact, together with the short-range intralayer ferromagnetic correlation revealed in isothermal magnetization measurements, suggests the possible presence of the Weyl state above $T_N$. These results demonstrate that in the $c$-axis magnetotransport of Mn(Bi$_{1-x}$Sb$_x$)$_2$Te$_4$, the spin scattering is dominant in the lightly electron-doped sample but overwhelmed by the chiral anomaly effect in the lightly hole-doped sample due to the presence of the Weyl state. These findings extend the understanding of the transport properties of Mn(Bi$_{1-x}$Sb$_x$)$_2$Te$_4$.**

*Email: zim1@psu.edu

## I. INTRODUCTION

MnBi$_2$Te$_4$ has recently been established as the first intrinsic magnetic topological insulator [1-3]. It is a layered van der Waals material composed of stacked Te-Bi-Te-Mn-Te-Bi-Te septuple layers (SL) along the crystallographic *c*-axis. Adjacent ferromagnetic (FM) Mn layers are coupled, forming an out-of-plane antiferromagnetic (AFM) order with $T_N$ = 25 K. Increasing the magnetic field along the *c*-axis causes MnBi$_2$Te$_4$ to undergo a spin-flop transition, manifested by the AFM to canted antiferromagnetic (CAFM) transition at $H_{c1}$ and the CAFM-to-FM transition at $H_{c2}$ [4,5]. At the same time, a nontrivial surface state is formed by inverted Bi and Te $p_z$ bands at the Γ point due to strong spin-orbital coupling (SOC). The combination of magnetism and nontrivial band topology in MnBi$_2$Te$_4$ leads to the realization of a quantum anomalous Hall insulator (QAHI) state, axion insulator, and layer Hall effect in the 2D thin layers [6-9]. Moreover, the quantized Hall effect with Chern number $C$ = 2 and 3 has also been demonstrated in MnBi$_2$Te$_4$ flakes [6,7,9-11]. Other exotic states, such as high-order topological insulator, Majorana hinge mode, and magnetic Skyrmion lattice, are also predicted to be realized in this material under certain conditions [12-14].

Additionally, MnBi$_2$Te$_4$ is also predicted to host an ideal type-II time-reversal symmetry (TRS) breaking Weyl semimetal (WSM) state when its AFM order is driven into FM order by a magnetic field parallel to the *c*-axis [1,2]. The recent theory further predicts that when the magnetic field is rotated away from the *c*-axis, the pair of Weyl points deviate from the $k_z$ axis, resulting in a type-I TRS breaking WSM until the Weyl points meet and annihilate each other at *H//ab*, turning MnBi$_2$Te$_4$ into a trivial FM insulator [15]. However, the field-driven WSM was not probed experimentally in pristine MnBi$_2$Te$_4$ since its Weyl nodes are not close to the Fermi energy ($E_F$). As such, chemical potential tuning is necessary to observe the predicted WSM in MnBi$_2$Te$_4$. Earlier work by Yan *et al*. and Chen *et al*. [16,17] have shown that the chemical potential of MnBi$_2$Te$_4$ can be tuned by Sb substitution for Bi. Recently, Lee *et al*. indeed observed experimental evidences for the predicted Weyl state by finely tuning the Sb concentration in Mn(Bi$_{1-x}$Sb$_x$)$_2$Te$_4$ [18]. They found that the system exhibits transport signatures of a TRS-breaking WSM state in its FM phase as the Sb concentration is tuned to ~26 %. In lightly hole-doped samples with $x \sim 0.26$, an electronic structure transition driven by the spin-flop transition is probed in the Hall resistivity and quantum oscillation measurements [18,19]. Such an electronic transition leads to a large negative *c*-axis longitudinal magnetoresistance (LMR) and a large intrinsic anomalous Hall effect, which provide strong support for the predicted FM WSM state [18].

All previous studies on Mn(Bi$_{1-x}$Sb$_x$)$_2$Te$_4$ have mostly focused on in-plane transport measurements [6,7,9-11,16,17,20]. Here, we report a comprehensive study of *c*-axis magnetotransport properties in Mn(Bi$_{1-x}$Sb$_x$)$_2$Te$_4$. Since the *c*-axis transport is sensitive to the interlayer spin scattering, systematic *c*-axis magnetotransport property measurements on Mn(Bi$_{1-x}$Sb$_x$)$_2$Te$_4$ would allow us to reveal how the spin scattering evolves with the chemical potential and how the spin scattering affects the *c*-axis transport of the FM WSM of the lightly hole-doped samples. We will focus on the comparison of *c*-axis magnetotransport properties between the lightly electron-doped and lightly hole-doped samples. Our prior work has shown that while the lightly electron-doped and lightly hole-doped samples share almost the same spin-flop transitions (i.e., the same $H_{c1}$ and $H_{c2}$) and similar carrier densities, transport signatures of the field-driven WSM were observed only in the lightly hole-doped samples. Through such a comparison, we anticipate advancing the understanding of the interplay between the spin scattering and the topological transport arising from the chiral anomaly effect of the WSM. Although we previously performed some *c*-axis magnetoresistivity measurements on several Mn(Bi$_{1-x}$Sb$_x$)$_2$Te$_4$ samples, those measurements were limited to low field ranges (≤9 T) [18]. All the measurements reported here were extended to 35 T, and the lightly electron-doped samples were not previously studied for their $c$-axis transport.

From our experiments, we observed several intriguing phenomena: (i) the spin scattering sensitively depends on the carrier type and carrier concentrations, which are determined by the chemical potential and tuned by the Sb concentration. The decrease of carrier density leads to significantly enhanced interlayer spin scattering, which results in a sharp increase in the $c$-axis resistivity $\rho_{zz}$ below the AFM ordering temperature $T_N$ for both the lightly electron- and hole-doped samples. (ii) The electronic anisotropy is significantly enhanced for the lightly hole-doped samples such that its paramagnetic states are characterized by striking incoherent transport behavior along the $c$-axis, manifested by a broad peak in the temperature dependence of $\rho_{zz}$ around 150 K. (iii) While both the lightly electron-doped and lightly hole-doped samples have the AFM states identical to that of $MnBi_2Te_4$ and show almost the same spin-flop transitions under magnetic fields [18], they exhibit distinct magnetotransport behavior along the $c$-axis: the lightly electron-doped samples display an extremely large spin-valve effect caused by the spin-flop transition, with the $c$-axis magnetoresistivity (MR) reaching ~ -95% at $H_{c2}$ and 6 K, whereas the lightly hole-doped samples exhibit large negative LMR which cannot be attributed to the spin-valve effect, but to the chiral anomaly effect of a WSM. These experimental observations indicate that when the Weyl nodes are close to the Fermi level, Weyl fermions dominate the $c$-axis transport and are not susceptible to spin scattering due to their relativistic effect, and the Weyl state extends to a wide temperature range above $T_N$ and likely exists even at zero magnetic field due to strong intralayer short-range FM correlations above $T_N$. These findings not only significantly extend the understanding of the dependence of the chiral anomaly and spin scattering on carrier type and concentration in $Mn(Bi_{1-x}Sb_x)_2Te_4$ but also provide an important framework for understanding magnetotransport properties of other relevant magnetic topological materials, MnBi$_{2n}$Te$_{3n+1}$ ($n$ = 2, 3 & 4) [21-24].

## II. Methods

The single crystals of Mn(Bi$_{1-x}$Sb$_x$)$_2$Te$_4$ were synthesized using the method reported in Ref. [18]. The phase purity of these single crystals was checked by X-ray diffraction. The sharp (00$l$) x-ray diffraction peaks demonstrate excellent crystallinity and the formation of the desired crystal structure in our single crystal samples. The composition analyses by energy-dispersive X-ray spectroscopy (EDS) show the actual Sb content $x$ slightly deviating from the nominal composition, as seen in our prior work [18]. In this article, we used the measured Sb content $x$ to label the samples used in this study.

The $c$-axis transport was measured using the standard four-probe method with the leads configured such that one current lead and one voltage lead were attached to each in-plane surface, as illustrated in the schematic of Fig. 1. In such a configuration, the applied current is expected to be aligned with the crystallographic $c$-axis. However, if the applied current is not exactly along the $c$-axis, the in-plane resistivity ($\rho_{xx}$) component can be involved in the measured out-of-plane resistivity $\rho_{zz}$. Although this situation likely occurs in our measurements, the sharp difference between the field dependences of the $c$-axis and in-plane magnetoresistivity clearly indicates that our measured $\rho_{zz}$ involves a small or negligible in-plane resistivity $\rho_{xx}$ component (see Supplementary Note 1 for detailed discussions). The low magnetic field transport measurements were performed using a commercial Physical Property Measurement System (PPMS, Quantum Design), while the high magnetic field transport measurements were carried out using the 35 T and 41.5 T resistive magnets at the NHMFL in Tallahassee. Field sweeps of the $c$-axis resistivity $\rho_{zz}$ were conducted for both positive and negative fields. The field dependence of $\rho_{zz}$ is obtained through symmetrizing the data collected at positive and negative fields, i.e., $\rho_{zz} = [\rho_{zz}(+\mu_oH) + \rho_{zz}(-\mu_oH)]/2$. High-field magnetization measurements were conducted at the Pulsed-Field Facility of the National High Magnetic Field Laboratory at Los Alamos National Laboratory using an extraction magnetometer in a short-pulse magnet. The absolute magnetization data were obtained by normalizing and calibrating from independent low-field magnetization data that were collected using SQUID magnetometer (Quantum Design).

## III. Results and Discussions

Figure 1(a) presents temperature-dependent normalized c-axis resistivity $[\rho_{zz}(T)/\rho_{zz}(T = 300 \text{ K})]$ on a logarithmic scale for representative $Mn(Bi_{1-x}Sb_x)_2Te_4$ samples with various Sb concentrations. All samples used for $\rho_{zz}$ measurements were cleaved from pieces with measured Hall resistivity $\rho_{yx}$, which was used to determine the carrier type and density. Table 1 summarizes the carrier type, carrier density, mobility, $H_{c1}$ & $H_{c2}$, and geometry information of all the samples used in this work (note that electron- and hole-doped samples are labeled with *E* and *H*, respectively). As seen in Fig. 1(a), $\rho_{zz}(T)$ strongly depends on carrier density. For the heavily doped samples with carrier density in the order of $10^{19}$-$10^{20}$ cm$^{-3}$, such as the *E1* and *H2* samples, $\rho_{zz}(T)$ shows metallic behavior in their PM states, similar to the in-plane resistivity $\rho_{xx}(T)$ [4,16,17]. However, when the carrier density is reduced to the order of $10^{18}$ cm$^{-3}$, $\rho_{zz}(T)$ maintains metallic behavior for the electron-doped sample *E*2, but displays remarkable non-metallic behavior for the hole-doped sample (*H*1), manifested by the broad peak around 150 K in the temperature dependence of $\rho_{zz}$. This suggests that when the hole Fermi pocket shrinks to a certain extent, it becomes highly anisotropic such that the c-axis transport becomes incoherent. A broad peak in the out-of-plane (i.e., c-axis) resistivity is a generic feature often seen in quasi-2D systems such as the ruthenate superconductor $Sr_2RuO_4$ [25]. This can be attributed to their quasi-2D electronic structures, as discussed as follows. For a layered anisotropic conductor with a quasi-2D electronic structure, the Fermi velocity along the *c*-axis is small. Thus, the *c*-axis hopping integral becomes very small. As a result, the mean free path along the *c*-axis ($l_c$) becomes smaller. When $l_c$ is shorter than the interplanar spacing, which usually occurs at high temperatures, band propagation along the *c*-axis is suppressed, thus resulting in charge confinement and incoherent charge transport between planes. In this case, the *c*-axis transport takes place via a diffusive or tunneling process. However, at low temperatures, the mean free path is increased due to the increased mean free time, such that band propagation along the *c*-axis can happen, thus leading to coherent, metallic-like transport.

From the $\rho_{zz}(T)$ data in Fig. 1(a), we also find that the interlayer spin scattering is also sensitive to the carrier density. As the carrier density is decreased from $10^{19}/10^{20}$ cm$^{-3}$ to $10^{18}$ cm$^{-3}$, $\rho_{zz}$ exhibits step-like jumps across $T_N$, leading to insulating-like behavior below $T_N$ (see the $\rho_{zz}$ data of samples *E*2 and *H*1 in Fig. 1(a)), which is sharply contrasted with the metallic behavior below $T_N$ in $\rho_{zz}$ and in-plane resistivity $\rho_{xx}$ of MnBi$_2$Te$_4$ [3,4,16]. These results clearly indicate that the interlayer spin scattering is significantly enhanced in the lightly electron-/hole-doped samples.

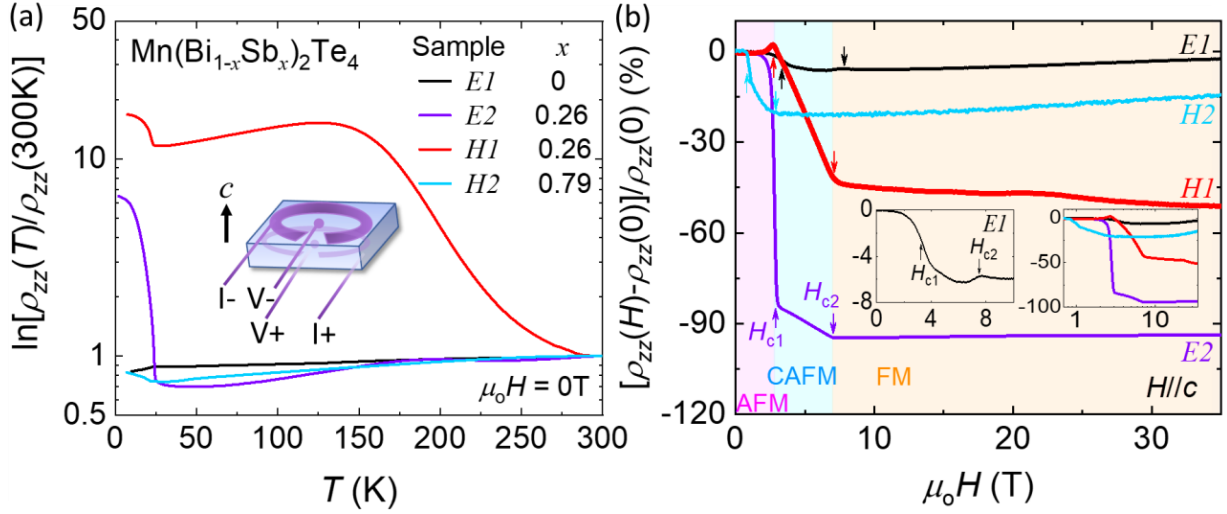

**Fig. 1.** Normalized $c$-axis resistivity of Mn(Bi$_{1-x}$Sb$_x$)$_2$Te$_4$ as the function of temperature for heavily electron-doped ($E1$), lightly electron-doped ($E2$), lightly hole-doped ($H1$), and heavily hole-doped ($H2$) samples. The schematic in (a) illustrates the experimental setup for the $c$-axis resistivity measurements. (b) Magnetic field dependence of $c$-axis magnetoresistivity for the representative heavily ($E1$, $H2$) and lightly ($E2$, $H1$) doped samples. The field is applied along the $c$-axis. The upward and downward arrows refer to the two magnetic transitions. The upward arrows indicate the AFM to CAFM transition at $H_{c1}$, and the downward arrow refers to the CAFM to FM transition at $H_{c2}$. The rose, blue, and orange regions refer to the AFM, CAFM, and FM phase regions for lightly doped samples $E2$ and $H1$. Left inset: zoomed-in $c$-axis magnetoresistivity for $E1$. Right inset: the $c$-axis magnetoresistivity plotted on the logarithmic scale of the magnetic field. The data of $E1$ is taken from [4] for comparison.

**Table. 1.** Information of the Mn(Bi$_{1-x}$Sb$_x$)$_2$Te$_4$ samples used in this study, including Sb content $x$, Neel temperature ($T_N$), critical field $H_{c1}$ (from AFM to CAFM) and $H_{c2}$ (from CAFM to FM) for $H//c$, carrier density, mobility, and the sample dimensions. The carrier density is estimated from the linear background of Hall resistivity $\rho_{yx}$ in the FM state at 2 K. The mobility is estimated in the PM phase at 75 K. All the samples used here for the $c$-axis transport measurements are cleaved from the pieces with known Hall carrier densities.

| Sample Label | Sb Content, $x$ (measured by EDS) | Sample Dimension ($w \times l \times t$) | Carrier Type | $T_N$ (K) | $H_{c1}$ (T) | $H_{c2}$ (T) | Carrier Density ($10^{20}$ cm$^{-3}$) | Hall Mobility (cm$^2$/Vs) |
|---|---|---|---|---|---|---|---|---|
| $E1$ | 0 | 0.75 × 0.88 × 0.25mm | electron-doped ($e$) | 25.0 | 3.6 | 7.7 | 1.3 | 58 |
| $E2$ | 0.26 | 0.83 × 0.91 × 0.13mm |  | 24.4 | 3.0 | 7.0 | 0.034 | 715 |
| $H1$ | 0.26 | 0.84 × 1.08 × 0.13mm | hole-doped ($h$) | 24.4 | 3.0 | 7.0 | 0.097 | 542 |
| $H2$ | 0.79 | 1.80 × 2.32 × 0.16mm |  | 21.4 | 0.9 | 3.1 | 0.55 | 23 |

Since interlayer spin scattering is dependent on spin polarization, the spin-flop transition is expected to suppress the interlayer spin scattering, thus resulting in a spin-valve effect. This was indeed observed in our prior work on MnBi$_2$Te$_4$ [4]. Its $c$-axis LMR resulting from the spin-valve effect is ~ -3.7% at $H_{c1}$ and 2 K (see the left inset to Fig. 1(b) for clarity). In the lightly electron-doped sample $E$2, we find its spin-valve effect at $H_{c1}$ is significantly enhanced, which is manifested by a step-like decrease at $H_{c1}$ in its $c$-axis LMR (LMR ~ -84% at $H_{c1}$ and 6 K, as denoted by the purple upward arrow in Fig. 1(b)). However, in the lightly hole-doped sample ($H$1), we did not observe such a sharp, step-like decrease expected for the spin-valve effect in its $c$-axis LMR at $H_{c1}$ though $H$1 shares almost the same spin-flop transition field with the lightly electron-doped sample (see Table 1 and Fig. 1(b)). Instead, the spin flop transition of sample $H$1 leads to only a gradual decrease in the $c$-axis LMR, and as the field is further increased above $H_{c2}$, its LMR continues to decrease. In our prior work [18], we attributed such a $c$-axis negative LMR probed in the lightly hole-doped samples to the chiral anomaly effect of a Weyl state since it exhibits a strong dependence on field orientation, sharply contrasted with the field orientation independent negative LMR caused by the spin-valve effect in the heavily electron/hole-doped samples. Our prior $c$-axis MR measurements on $H$1 were made only up to 9 T, while our current measurements were extended to 35 T. From $H_{c2}$ to 35 T, the LMR of $H$1 further decreases by ~8% (Fig. 1(b)). Such behavior is in stark contrast with other samples whose $c$-axis LMR either displays a slight upturn above $H_{c2}$ (for $E$1 & $H$2) or almost remains constant above $H_{c2}$ (for $E$2). This contrast can be seen clearly in the right inset of Fig. 1(b), which is plotted on the logarithmic scale of the magnetic field. Such an unusual field dependence of LMR of sample $H$1 is suggestive of the expected chiral anomaly effect.

To further demonstrate that the lightly hole-doped sample exhibits the chiral anomaly effect while the lightly electron-doped sample shows only the spin-valve effect, we also measured the dependence of $c$-axis MR on different field orientations under high magnetic fields for samples $E2$, $H1$, and $H2$. As shown in Fig. 2(a), the $c$-axis MR of sample $H1$ at 6 K displays a strong angular dependence. Its magnitude of negative MR gradually decreases as the field tilt angle $\theta$ relative to the $c$-axis is increased (see the inset to Fig. 2(c) for the experimental setup), and the sign of MR switches from negative to positive as $\theta$ is increased above 49°. Such a strong angular dependence of the $c$-axis MR was also reproduced in another two lightly hole-doped samples (see Supplementary Note 2). In contrast, in sample $E2$ which shows a strong spin-valve effect, we find its $c$-axis MR exhibits only a weak angular dependence above $H_{c2}$ (Fig. 2(c)). Another remarkable feature seen in this sample is that its MR is nearly saturated above $H_{c2}$ (note that $H_{c2}$ slightly increases from $\theta = 0°$ to 90° as denoted by the dotted line), which agrees well with the spin-valve picture discussed above. For the heavily hole-doped sample $H2$, which exhibits only a weak spin-valve effect, its $c$-axis MR also exhibits very weak angular dependence (Fig. 2(e)). The sharp contrast in the angular dependence of the $c$-axis MR between $H1$, $E2$, and $H2$ (Figs. 2(a), 2(c) & 2(e)) clearly indicates that the lightly hole-doped sample has a distinct transport mechanism in its CAFM and FM phases compared with other samples. According to the above discussions and our prior work [18], the chiral anomaly effect of the field-driven FM Weyl state can account for all the anomalous $c$-axis magnetotransport behavior under high magnetic fields for the lightly hole-doped sample $H1$. In other words, as the Weyl nodes are present near $E_F$ in sample $H1$, the chiral anomaly effect overwhelms the spin-valve effect. Such a transport mechanism is further corroborated by the measurements carried out at 25 K (which is slightly above $T_N = 24.4$ K). Although the long-range AFM order is suppressed at 25 K, a short-range AFM order should survive. High magnetic fields are expected to drive it to a forced FM phase in both *H*1 and *E*2 samples. As such, we naturally expect a suppressed spin-valve effect at 25 K in the lightly electron-doped sample *E*2 but a suppressed chiral anomaly effect at 25 K in the lightly hole-doped sample *H*1. The data presented in Figs. 2(b) & 2(d) are in good agreement with such anticipation. Furthermore, we find that the heavily hole-doped sample *H*2 displays only small, field orientation-independent negative MR at 25 K (Fig. 2(f)), which can be ascribed to the gradual suppression of spin scattering by the magnetic field. These data again indicate that the topological quantum transport associated with the Weyl state is accessible only in the lightly hole-doped samples and Weyl fermions are insusceptible to interlayer spin scattering due to the linear dispersion of Weyl bands.

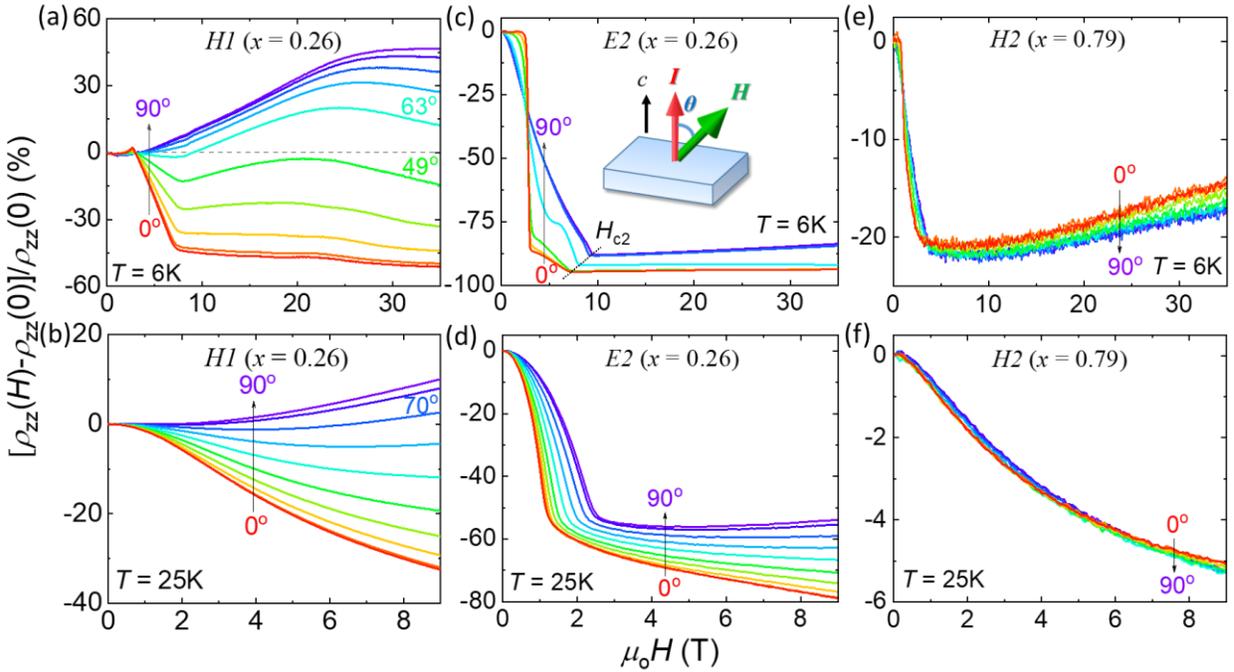

**Fig. 2.** *c*-axis magnetoresistivity MR = $[\rho_{zz}(H) - \rho_{zz}(0)]/\rho_{zz}(0)$ under various field orientations of the lightly hole- (*H1*), lightly electron- (*E2*), and heavily hole-doped (*H2*) samples at 6 K (a,c,e) and 25 K (b,d,f). The schematic in (c) illustrates the experimental setup for the *c*-axis MR angular dependence measurements.

From the temperature-dependent measurements of the LMR of sample *H*1, we also find that its chiral anomaly effect extends to a wide temperature range above $T_N$. In Fig. 3(a), we present the *c*-axis LMR data of this sample at various temperatures, which show large negative values even as the temperature is increased up to 90 K. When the field is rotated to the in-plane direction (i.e., $H \perp I$), the transverse MR becomes positive (Fig. S1). These observations imply that the negative LMR above $T_N$ of sample *H*1 is not due to suppression of spin scattering since the suppression of spin scattering by magnetic fields is weakly dependent on field orientation, as revealed above. Such negative LMR should also not be associated with the anomalous velocity induced by nonzero Berry curvature [26] or the Zeeman effect [27]. Negative LMR caused by these two mechanisms has been demonstrated in several other topological insulators. The LMR due to the anomalous velocity mechanism is usually very small (*e.g.,* see [26]) and independent of the current direction. Our observed negative LMR behavior in sample *H*1 clearly does not seem to fit into this mechanism since its negative LMR is very large (*e.g.,* LMR= -31.5% even at 24 K and 9 T), and we did not observe negative LMR in the in-plane magnetotransport measurements with *H*//*I* and *I*//*ab*-plane (see supplementary Fig. S2). The Zeeman effect-induced negative LMR occurs on a barely percolating current path formed in the disordered bulk materials [27]. This mechanism is applicable in sufficiently bulk-insulating materials with the chemical potential inside the gap. Given that the current paths are greatly affected by the magnetic fields due to the Zeeman effect that leads to an essentially isotropic negative MR, we expect to observe negative MR regardless of the experimental configuration; negative MR should appear for both longitudinal (*H*//*I*) or transverse ($H \perp I$) configurations [27]. However, in sample *H*1, the sign of the *c*-axis MR changed from negative to positive when the tilted angle of the magnetic field is greater than 49° (Fig. 2(a)). In addition, if the Zeeman effect-mechanism was the origin of the negative LMR, negative MR would also be expected for the in-plane MR measurements, which is contradictory to our observation of positive in-plane LMR (supplementary Fig. S3).

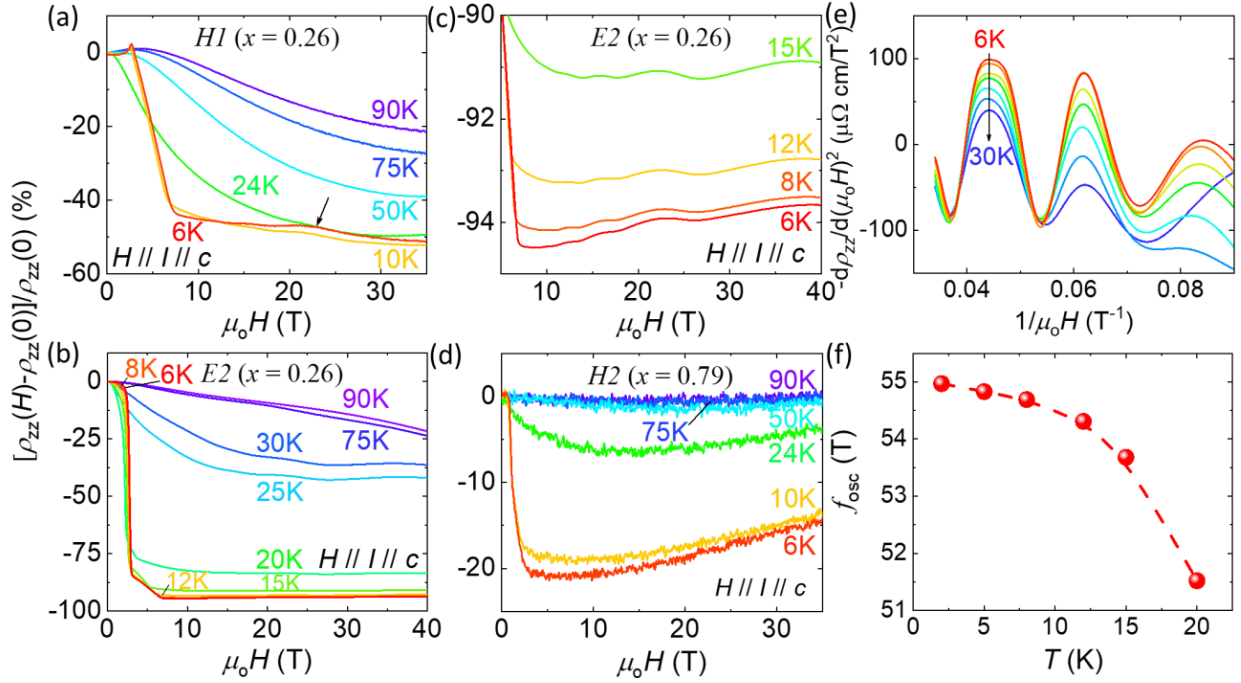

**Fig. 3.** *c*-axis longitudinal magnetoresistivity MR = [$\rho_{zz}(H) - \rho_{zz}(0)$]/$\rho_{zz}(0)$ at various temperatures for (a) the lightly hole- (*H1*), (b) lightly electron- (*E2*), and (d) heavily hole-doped (*H2*) samples. The field is applied along the *c*-axis. (c) Zoomed-in data of panel (b), which shows the SdH oscillations of sample *E2*. (e) The second derivative of the *c*-axis MR with respect to field *H* for sample *E2* as a function of the inverse of the field at various temperatures. (f) Oscillation frequency ($f_{osc}$) as a function of temperature for sample *E2*.

After excluding the anomalous velocity and Zeeman effect as the origin of the negative LMR seen in the lightly hole-doped sample above $T_N$, let's discuss the possible origin of the chiral anomaly effect. In general, a WSM is expected to exhibit a $B^2$-dependence of magnetoconductivity when its transport is dominated by the chiral anomaly effect under parallel electric and magnetic fields [28,29]. In this case, its total conductivity can be expressed as $\sigma = \sigma_o(1 + C_w B^2)$, where $\sigma_o$ is the normal conductivity and $\sigma_o C_w B^2$ is the chiral anomaly contribution. If the system involves weak antilocalization (which is the case for our lightly hole-doped sample *H*1), the normal conductivity should be corrected to $\sigma_o + \alpha \sqrt{B}$ [30]. Thus, the total conductivity is modified to $\sigma = (\sigma_o + \alpha \sqrt{B})(1 + C_w B^2)$. Using this equation, we can nicely fit the *c*-axis longitudinal magnetoconductivity data of sample $H$1 at 50 K, 75 K, and 90 K in moderate field ranges (0-10 T for 50 K, 0-13 T for 75 K, and 0-17.5 T for 90 K; the fits for the data measured at 30 K, 40 K, and 60 K are presented in Supplementary Fig. S4), as shown in Fig. 4. The fitted $C_w$ is hardly temperature-dependent for these three temperatures (see the inset to Fig. 4). These fitted results provide strong support for the above argument that the PM states of the lightly hole-doped sample likely host a WSM state. The high field deviation of magnetoconductivity from $\sigma = (\sigma_o + \alpha \sqrt{B})(1 + C_w B^2)$ is because the classic magnetoresistance ($\propto B^2$), which becomes much larger at high magnetic fields, was not taken into account in our fits. However, the fit for the data at 24 K (slightly below $T_N$) is limited to a much smaller field range, and the extracted $C_w$ is also much greater than those obtained at high temperatures, so we did not include this data point in the inset of Fig. 4. This can be attributed to the fact that the negative MR due to the suppression of spin scattering is enhanced near $T_N$, and this component is not considered in the fit. Additionally, it should also be pointed out that the magnetoconductivity data of 6 K and 10 K cannot be fitted with the above equation above $H_{c2}$ either. The presence of quantum oscillations may account for this deviation; an SdH oscillation peak near 22 T was indeed observed at 6 K and 10 K, as denoted by the arrow in Fig. 3(a).

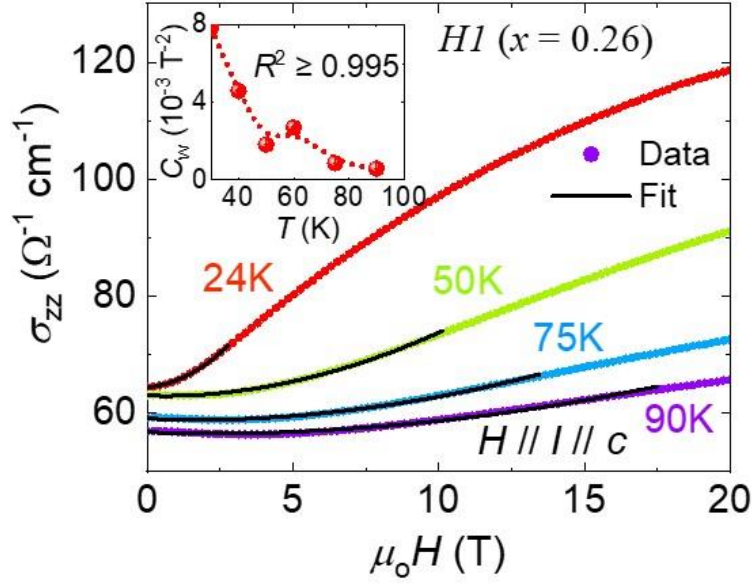

**Fig. 4.** Field dependence of magnetoconductivity of sample $H$1 at various temperatures. The black curves represent the fits to the equation $\sigma = (\sigma_o + \alpha \sqrt{B})(1 + C_w B^2)$. The inset shows the temperature dependence of $C_w$ extracted from the fitting.

The fact that the fit of magnetoconductivity to $\sigma = (\sigma_o + \alpha \sqrt{B})(1 + C_w B^2)$ can extend to zero magnetic field (Fig. 4) implies that the WSM might be present even at zero field at temperatures above $T_N$ for the lightly hole-doped sample. For the material system studied in this work, its WSM must require broken time-reversal symmetry (TRS), as mentioned above. A paramagnetic (PM) state is not generally expected to break TRS. Nevertheless, broken TRS can be present if the PM state features FM fluctuations or static short-range FM order [31]. We note that spin-fluctuations induced Weyl state has been observed in EuCd$_2$As$_2$ [32]. For MnBi$_2$Te$_4$, prior DFT calculations and electron spin resonance (ESR) experiments at $T > T_N$ have proven that its PM state is characterized by strong intralayer FM correlations [3,33]. To find if our lightly hole-doped samples possess strong intralayer FM correlations, we performed isothermal magnetization measurements on a lightly hole-doped sample ($H$1-$d$) up to 35 T in a wide temperature range (1.5 K - 90 K). Figure 5 presents the measured data. The data at 1.5 K agrees with our previously published data [18] and reveals a remarkable spin-flop transition from the AFM to the canted AFM state at $H_{c1} = 3$ T and then to the FM state at $H_{c2} = 7$ T. When the field is above $H_{c2}$, the magnetization is saturated, with a saturated moment of $M_s \sim 3.52$ $\mu_B$/f.u. Due to the existence of antisite defects, the magnetization gradually increases above 20 T to 4.3 $\mu_B$/f.u at 35 T [34]. At $T_N$ (~ 24 K), while a clear spin-flop transition diminishes, the magnetization shows a striking sublinear increase with increasing field. Such an FM polarization behavior extends to high temperatures and is still discernable even at 90 K, suggesting that strong short-range intralayer FM correlation is present above $T_N$, consistent with the previously reported ESR experimental results and DFT calculations [3,33]. In other words, the system is nearly FM in a wide temperature window close to $T_N$ though the FM Mn layers are aligned antiferromagnetically below $T_N$. The interlayer AFM coupling above $T_N$ should be much weaker than the intralayer FM coupling. This is generally expected for layered systems with *A*-type AFM orders [35,36]. Under high magnetic fields, such a nearly FM state is driven to a forced FM state. Therefore, the Weyl state can be present in such a PM state under magnetic fields or even at zero field from the theoretical point of view. Of course, spectroscopy experiments (i.e., ARPES measurements) are needed to find direct evidence for the zero-field Weyl state, which is beyond the scope of this work.

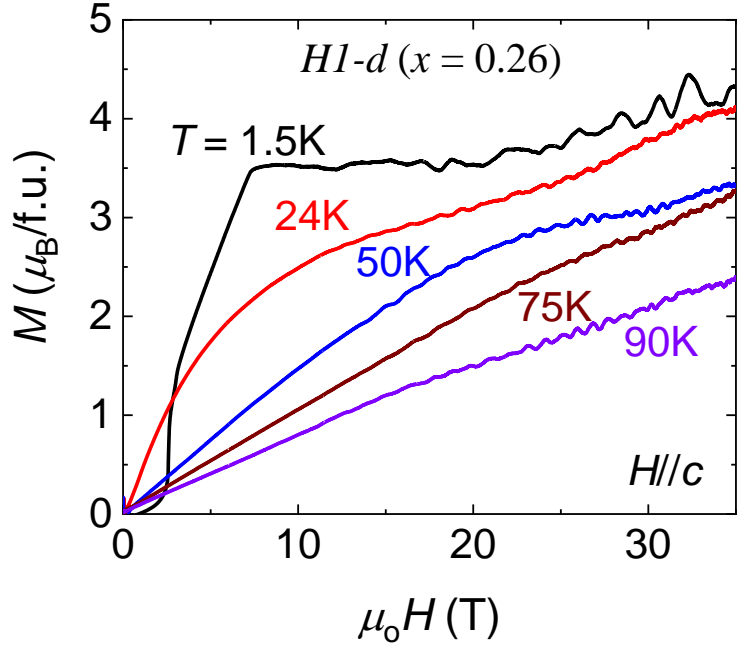

**Fig. 5.** High-field isothermal magnetization data of lightly hole-doped sample Mn(Bi$_{1-x}$Sb$_x$)$_2$Te$_4$ with $x = 0.26$ at various temperatures and $H//c$-axis.

For the lightly electron-doped sample ($E$2), we did not find any features consistent with a WSM in its PM state. Its negative LMR, which arises from the spin-valve effect, is also temperature-dependent (Fig. 3(b)); its magnitude of LMR remarkably decreases as the temperature is increased above $T_N$ (~24 K), with its field dependence distinct from the field dependence of the negative LMR caused by the chiral anomaly effect in sample $H$1. In this sample, we also observed Shubnikov–de Haas (SdH) oscillations above $H_{c2}$ where the FM phase occurs; this can be seen clearly from the zoomed-in data in Fig. 3(c). By taking the second derivative, we extracted its oscillatory component of MR (see Fig. 3(e)). Through Fourier transformation analyses, we obtained the quantum oscillation frequencies, which are found to be temperature-dependent (Fig. 3(f)). The oscillation frequency increases with the decrease of temperature, consistent with the prior results extracted from the SdH oscillations of in-plane TMR in the lightly electron-doped sample and suggests a strong coupling between the electronic structure and magnetism, as discussed in prior reports [18,19]. In the heavily hole-doped sample $H2$, which exhibits only weak spin-valve behavior, its LMR is almost fully suppressed above $T_N$, as shown in Fig. 3(d).

## IV. Conclusions

In summary, we have systematically studied the $c$-axis transport properties of Mn(Bi$_{1-x}$Sb$_x$)$_2$Te$_4$. We find that the interlayer spin scattering is sensitive to carrier density in this system. When the carrier density is reduced from the heavy ($10^{19}$ - $10^{20}$ cm$^{-3}$) to the light doping level (~$10^{18}$ cm$^{-3}$), the interlayer spin scattering in the AFM state is significantly enhanced for both lightly electron- and hole-doped samples, leading to a step-like increase in $\rho_{zz}(T)$. Although the lightly electron- and hole-doped samples have comparable carrier densities and share similar spin-flop transitions, their $c$-axis transport shows a distinct response to the spin-flop transition. The lightly electron-doped sample exhibits a giant spin-valve effect upon the spin-flop transition, while the lightly hole-doped sample displays a remarkable negative LMR consistent with the chiral anomaly, and its LMR continues to decrease with the increasing field above $H_{c2}$, suggesting Weyl fermions are insusceptible to spin scattering. Moreover, the Weyl state of the lightly hole-doped sample is found to extend to the PM state due to the strong intralayer FM correlations. This is evidenced by the observation that the $c$-axis magnetoconductivity of the lightly hole-doped sample follows $B^2$ dependence in a wide temperature range above $T_N$ as well as the FM-like magnetic polarization in the PM state. Given that the Weyl state is also predicted to be present in other relevant topological materials MnBi$_{2n}$Te$_{3n+1}$, our findings provide an important framework to search for the predicted Weyl states in those materials.

**Acknowledgment**

The study is based upon research conducted at The Pennsylvania State University Two-Dimensional Crystal Consortium–Materials Innovation Platform (2DCC-MIP), which is supported by NSF Cooperative Agreement No. DMR-2039351. Z.Q.M. and R.R. acknowledge the support from NSF under Grant No. DMR 2211327. The work at the National High Magnetic Field Laboratory is supported by the NSF Cooperative Agreement No. DMR1157490 and the State of Florida.## References

[1] D. Zhang, M. Shi, T. Zhu, D. Xing, H. Zhang, and J. Wang, *Physical Review Letters* **122**, 206401 (2019).
[2] J. Li, Y. Li, S. Du, Z. Wang, B.-L. Gu, S.-C. Zhang, K. He, W. Duan, and Y. Xu, *Science Advances* **5**, eaaw5685 (2019).
[3] M. M. Otrokov *et al.*, *Nature* **576**, 416 (2019).
[4] S. H. Lee *et al.*, *Physical Review Research* **1**, 012011 (2019).
[5] J. Q. Yan *et al.*, *Phys. Rev. Mater.* **3**, 8, 064202 (2019).
[6] Y. Deng, Y. Yu, M. Z. Shi, Z. Guo, Z. Xu, J. Wang, X. H. Chen, and Y. Zhang, *Science*, eaax8156 (2020).
[7] C. Liu *et al.*, *Nat Mater* **19**, 522 (2020).
[8] A. Gao *et al.*, *Nature* **595**, 521 (2021).
[9] J. Cai *et al.*, *arXiv:* , 2107.04626 (2021).
[10] J. Ge, Y. Liu, J. Li, H. Li, T. Luo, Y. Wu, Y. Xu, and J. Wang, *National Science Review* (2020).
[11] C. Liu *et al.*, *Nature Communications* **12**, 4647 (2021).
[12] R.-X. Zhang, F. Wu, and S. Das Sarma, *Physical Review Letters* **124**, 136407 (2020).
[13] Y. Peng and Y. Xu, *Physical Review B* **99**, 195431 (2019).
[14] B. Li *et al.*, *Physical Review Letters* **124**, 167204 (2020).
[15] P. Wang, J. Ge, J. Li, Y. Liu, Y. Xu, and J. Wang, *The Innovation* **2**, 100098 (2021).
[16] J. Q. Yan, S. Okamoto, M. A. McGuire, A. F. May, R. J. McQueeney, and B. C. Sales, *Physical Review B* **100**, 104409 (2019).
[17] B. Chen *et al.*, *Nature Communications* **10**, 4469 (2019).
[18] S. H. Lee *et al.*, *Physical Review X* **11**, 031032 (2021).
[19] Q. Jiang *et al.*, *Physical Review B* **103**, 205111 (2021).
[20] Y. Chen *et al.*, *Phys. Rev. Mater.* **4**, 064411 (2020).
[21] J. Q. Yan, Y. H. Liu, D. S. Parker, Y. Wu, A. A. Aczel, M. Matsuda, M. A. McGuire, and B. C. Sales, *Phys. Rev. Mater.* **4**, 054202 (2020).
[22] L. Ding, C. Hu, F. Ye, E. Feng, N. Ni, and H. Cao, *Physical Review B* **101**, 020412 (2020).
[23] J. Wu *et al.*, *Science Advances* **5**, eaax9989 (2019).
[24] I. I. Klimovskikh *et al.*, *npj Quantum Materials* **5**, 54 (2020).
[25] N. E. Hussey, A. P. Mackenzie, J. R. Cooper, Y. Maeno, S. Nishizaki, and T. Fujita, *Physical Review B* **57**, 5505 (1998).20